\documentclass[12pt]{article}
\usepackage{amscd,amsfonts,amssymb,amsmath,latexsym,array,hhline}
\usepackage[dvips]{graphics}
\makeatletter
\@addtoreset{equation}{section}
\renewcommand{\@begintheorem}[2]{\begin{trivlist}\it
\item[\hspace{\labelsep}{\bf #1\ #2.}]}
\renewcommand{\@opargbegintheorem}[3]{\begin{trivlist}\it
\item[\hspace{\labelsep}{\bf #1\ #2\ (#3).}]}
\renewcommand{\@endtheorem}{\end{trivlist}}
\renewcommand{\@cite}[2]{[{#1\if@tempswa ; #2\fi}]}
\makeatother

\newcommand{\paragr}{\hspace{6mm}}
\newtheorem{theorem}{\paragr Theorem}[section]

\newtheorem{corollary}[theorem]{\paragr Corollary}

\newtheorem{example}[theorem]{\paragr Example}

\newtheorem{lemma}[theorem]{\paragr Lemma}

\newtheorem{proposition}[theorem]{\paragr Proposition}

\newenvironment{mitemize}%
{\begin{list}{$\bullet$}{
\leftmargin=32pt
\rightmargin=0pt
\labelsep=5pt
\labelwidth=20pt
\itemindent=0pt
\topsep=5pt plus 2pt minus 4pt
\partopsep=2pt plus 1pt minus 1pt
\parsep=0pt
\itemsep=0pt}}%
{\end{list}}

\begin{document}
\vspace*{5mm}
\begin{center}\bf\large
CONCAVE DISTORTION SEMIGROUPS
\end{center}

\begin{center}\itshape\bfseries\large
Alexander Cherny$^*$\qquad Damir Filipovi\'c$\,^{**}$
\end{center}

\begin{center}
\textit{$^*$Department of Probability Theory}\\
\textit{Faculty of Mechanics and Mathematics}\\
\textit{Moscow State University}\\
\textit{119992 Moscow Russia}\\
\texttt{E-mail: alexander.cherny@gmail.com}
\end{center}

\begin{center}
\textit{$^{**}$Vienna Institute of Finance}\\
\textit{University of Vienna and Vienna University of Economics and Business Administration}\\
\textit{1090 Vienna}\\
\texttt{E-mail: Damir.Filipovic@wu-wien.ac.at}
\end{center}

\begin{center}
January 11, 2008
\end{center}

\begin{abstract}
\textbf{Abstract.}
The problem behind this paper is the proper measurement of the degree of
quality/acceptability/distance to arbitrage of trades.
We are narrowing the class of coherent acceptability indices introduced
by Cherny and Madan~\cite{CM071} by imposing an additional mathematical property.
For this, we introduce the notion of a \textit{concave distortion semigroup}
as a family $(\Psi_t)_{t\ge0}$ of concave increasing functions
$[0,1]\to[0,1]$ satisfying the semigroup property
$$
\Psi_s\circ\Psi_t=\Psi_{s+t},\quad s,t\ge0.
$$
The goal of the paper is the investigation of these semigroups with regard to the
following aspects:
\begin{mitemize}
\item representation of distortion semigroups;
\item properties of distortion semigroups desirable from the economical or
mathematical perspective;
\item determining which concave distortions belong to some distortion semigroup.
\end{mitemize}

\textbf{Key words and phrases:}
Coherent acceptability index,
concave distortion semigroup,
generator of a distortion semigroup,
Wang transform,
Weighted V@R.
\end{abstract}

\section{Introduction}
\label{I}

\textbf{1. Two problems.}
Let $X$ be a random variable modelling the discounted profit and loss produced
by some portfolio over the unit time period, i.e. $X=W_1-W_0$, where $W_0$ and $W_1$ denote the
initial and terminal value of the portfolio, respectively.
It is important for various purposes to measure the degree of quality of~$X$.
This is needed, in particular, for measuring the degree of market efficiency
and for portfolio optimization; measures of performance can also be applied to
pricing through the No Good Deals condition.
The classical measures of performance are the Sharpe ratio
$\text{SR}(X)=\mathsf{E}X/\sigma(X)$, where $\sigma$ is the standard deviation,
the Risk-Adjusted Return on Capital
$\text{RAROC}(X)=\mathsf{E}X/\text{V@R}(X)$, where V@R is the Value at Risk,
and the Gain-Loss Ratio
$\text{GLR}(X)=\mathsf{E}(X)/\mathsf{E}X^-$, where $X^-=\max\{-X,0\}$.
Each of these measures has its advantages and disadvantages.
Cherny and Madan~\cite{CM071} addressed the following problem: which
properties should a proper measure of performance satisfy?
They introduced the notion of a \textit{coherent acceptability index} as a
map $\alpha:L^\infty\to[0,\infty]$ (we have a fixed probability space
$(\Omega,\mathcal{F},\mathsf{P})$) satisfying the axioms:
\begin{mitemize}
\item (Quasi-concavity) if $\alpha(X)\ge z$ and $\alpha(Y)\ge z$, then $\alpha(X+Y)\ge z$;
\item (Monotonicity) if $X\le Y$, then $\alpha(X)\le\alpha(Y)$;
\item (Scale invariance) $\alpha(\lambda X)=\alpha(X)$ for
$\lambda\in(0,\infty)$;
\item (Fatou property) if $|X_n|\le1$, $X_n$ converge to $X$ in probability,
and $\alpha(X_n)\ge z$, then $\alpha(X)\ge z$.
\end{mitemize}
Let us remark that of the three performance measures described above (SR,
RAROC, and GLR) only the last one satisfies these axioms.
However, GLR is not the best representative of coherent indices; there exist
indices with better properties that are described below.

The representation theorem provided in~\cite{CM071} states that a map $\alpha$
is a coherent acceptability index if and only if there exists a family
$(\mathcal{D}_t)_{t\ge0}$, where each $\mathcal{D}_t$ consists of probability
measures absolutely continuous with respect to~$\mathsf{P}$, such that
$\mathcal{D}_s\subseteq\mathcal{D}_t$ for $s\le t$ and
\begin{equation}
\label{i1}
\alpha(X)=\sup\Bigl\{t\ge0:\inf_{\mathsf{Q}\in\mathcal{D}_t}
\mathsf{E}_\mathsf{Q}X\ge0\Bigr\},
\end{equation}
where $\sup\emptyset=\infty$.
This representation shows that coherent indices are closely linked to coherent
risk measures introduced by Artzner, Delbaen, Eber, and Heath~\cite{ADEH97},
\cite{ADEH99}. Namely, $\alpha$~is a coherent index if and only if there
exists a family $(\rho_t)_{t\ge0}$ of coherent risk measures such that $\rho_s\le\rho_t$ for $s\le t$ and
$$
\alpha(X)=\sup\{t\ge0:\rho_t(X)\le0\}.
$$

The major disadvantage of coherent indices is that this class is very wide and
at first it is not clear which particular index should be chosen in applications.
Indeed, the class of coherent risks is already very wide, and an index is linked
to a whole family of coherent risks.
An analytically convenient subclass of coherent indices is the class of
\textit{Weighted V@R acceptability indices} (see~\cite[Subsect.~3.6]{CM071})
denoted as AIW and defined as
\begin{equation}
\label{i2}
\alpha(X)=\sup\Bigl\{t\ge0:\int_\mathbb{R}xd(\Psi_t(D_X(x)))\ge0\Bigr\},
\end{equation}
where $D_X$ is the distribution function of~$X$ and
$(\Psi_t)_{t\ge0}$ is a family of \textit{concave distortions}
increasing in~$t$, i.e each $\Psi_t$ is a concave increasing
function $[0,1]\to[0,1]$ with $\Psi_t(0)=0$, $\Psi_t(1)=1$ and $\Psi_t(x)$ is
increasing in~$t$ for any $x\in[0,1]$.
Each map $X\mapsto-\int_{\mathbb{R}}xd(\Psi_t(D_X(x)))$ is a coherent risk
measure from the \textit{Weighted V@R} class known also as
\textit{distorted probability measures} and \textit{spectral risk measures}.
These date back to Yaari~\cite{Y87} and Denneberg~\cite{D90};
for more details on this class, see~\cite{A02}, \cite{C06}, and \cite[Sect.~4.6]{FS04}.

However, the AIW class is still too wide.
There are some economically desirable properties that enable one to slightly
narrow the class of reasonable indices, but this reduction is actually negligible.
The first basic problem we address is:
\textit{To select a subclass of AIW that is essentially narrower than AIW by
imposing some mathematically desirable condition}.

The second problem under consideration also deals with the problem of the
selection of appropriate indices and is as follows. Assuming that market
participants are maximizing some coherent index~$\alpha$ from the AIW class,
one can recover by comparing the physical measure (estimated from the
historical data) and the risk-neutral measure (estimated from option prices)
the concave distortion $\Psi_{t_*}$ corresponding to $t_*=\alpha(X_*)$, where $X_*$
is the portfolio of the representative agent; for details, see~\cite{CM073}.
In other words, from the data one might recover the form of~$\Psi_t$ for
some particular~$t_*$. Then one might try to recover the family
$(\Psi_t)_{t\ge0}$ by the function $\Psi_{t_*}$. Clearly, there exists a huge
variety of such families. The problem we want to address is: \textit{Does
there exist a ``canonical'' family of concave distortions $(\Psi_t)_{t\ge0}$
with a given $\Psi_{t_*}$?}

\medskip
\textbf{2. Goal of the paper.}
Let us make the following observation: if $\Psi$ and $\widetilde\Psi$ are two
concave distortions, then its composition $\Psi\circ\widetilde\Psi$ is again a
concave distortion and it dominates both~$\Psi$ and~$\widetilde\Psi$.
A mathematically elegant property of an increasing family of concave
distortions $(\Psi_t)_{t\ge0}$ is the semigroup property
$$
\Psi_s\circ\Psi_t=\Psi_{s+t},\quad s,t\ge0.
$$

The goal of the paper is to investigate families of concave distortions with
the semigroup property. We will call them \textit{concave distortion
semigroups} or simply \textit{distortion semigroups} (the word
``distortion'' will also be skipped at some places as we are not considering
any other semigroups).
For this class, the above two problems transform into the following ones:
\begin{mitemize}
\item[\bf I.] \textit{How can distortion semigroups be represented?}
\item[\bf II] \textit{For which concave distortions $\Psi$ does there exist a
distortion semigroup $(\Psi_t)_{t\ge0}$ such that $\Psi_1=\Psi$?}
\end{mitemize}

\medskip
\textbf{3. Examples.}
Let us first give several examples showing that the class of distortion
semigroups includes some nice representatives.

It is easy to see that the family
\begin{equation}
\label{i3}
\Psi_t(x)=(e^tx)\wedge1,\quad t\ge0,\;x\in[0,1]
\end{equation}
belongs to this class. For $X$ with a continuous distribution, we have
$$
\int_\mathbb{R}xd(\Psi_t(D_X(x)))=\mathsf{E}(X\mid X\le q_{e^{-t}}(X)),
$$
where $q_\lambda(X)$ is a $\lambda$-quantile of~$X$.
Hence, the corresponding AIW index has a nice representation
$$
\alpha(X)=-\ln(\inf\{\lambda\in(0,1]:
\mathsf{E}(X\mid X\le q_\lambda(X))\ge0\}).
$$
This index coincides up to a monotone transformation with the \textit{CV@R}
index from~\cite{CM071}.\footnote{Note that if $\alpha$ is a coherent index
and $\varphi:\mathbb{R}_+\to\mathbb{R}_+$ is a continuous strictly increasing
function, then $\varphi\circ\alpha$ is again a coherent index, which we call
a monotone transformation of~$\alpha$.}

Another example of a distortion semigroup is
\begin{equation}
\label{i4}
\Psi_t(x)=1-(1-x)^{e^t},\quad t\ge0,\;x\in[0,1].
\end{equation}
If $e^t$ is integer, then
$$
\int_\mathbb{R}xd(\Psi_t(D_X(x)))=\mathsf{E}\min\{X_1,\dots,X_{e^t}\},
$$
where $X_1,X_2,\dots$ are independent copies of~$X$.
Thus, $\alpha(X)$ admits an informal interpretation as the largest number~$t$
such that the expectation of the minimum of~$e^t$ independent draws of~$X$ is
positive. This index coincides up to a monotone transformation with the
\textit{AIMIN} index from~\cite{CM071}.

A further example is
\begin{equation}
\label{i5}
\Psi_t(x)=x^{e^{-t}},\quad t\ge0,\;x\in[0,1].
\end{equation}
The corresponding index coincides up to a monotone transformation with the
\textit{AIMAX} index from~\cite{CM071}.

One more example is the \textit{Wang transform} introduced in~\cite{W00} and defined as
\begin{equation}
\label{i6}
\Psi_t(x)=\Phi(\Phi^{-1}(x)+t),\quad t\ge0,\;x\in[0,1],
\end{equation}
where $\Phi$ is the distribution function of the standard normal distribution.
Clearly, $(\Psi_t)_{t\ge0}$ has the semigroup property; a direct differentiation shows
that each $\Psi_t$ is concave.

\medskip
\textbf{4. Results.}
The representation of distortion semigroups is provided by Theorem~\ref{R1}.
It states that $(\Psi_t)_{t\ge0}$ is a distortion semigroup if and only if
there exists a concave function $G:(0,1)\to(0,\infty)$ such that
$$
\Psi_t(x)=\inf\Bigl\{y\in[x,1]:\int_x^y\frac{1}{G(s)}\,ds=t\Bigr\},
\quad t\ge0,\;x\in[0,1].
$$
The function $G$ is uniquely determined by the family $(\Psi_t)_{t\ge0}$ via
the relation
$$
G(x)=\lim_{t\downarrow0}\frac{\Psi_t(x)-x}{t},\quad x\in(0,1),
$$
and $G(0)=G(1):=0$. We call $G$ the \textit{generator} of the semigroup $(\Psi_t)_{t\ge0}$.

In order to compare different distortion semigroups and to identify
which ones are better, we investigate in Section~\ref{P} how some economically and mathematically desirable properties
 are expressed in terms of generators.
As an example, one of such properties is the strict quasi-concavity, which
guarantees the uniqueness of a solution of optimization problems based on the
corresponding index.

Next we address the second problem mentioned above.
Theorem~\ref{L1} states that for a concave distortion~$\Psi$ there exists at
most one distortion semigroup $(\Psi_t)_{t\ge0}$ with $\Psi_1=\Psi$
(under the additional assumption on the left derivative $\Psi'_-(1)>0$, which is in fact economically
reasonable).
It also provides an approximation procedure for finding its generator.
However, such a semigroup does not exist for all the concave distortions~$\Psi$
as shown by Example~\ref{L2}.
The problem of describing the class of concave distortions, for which there
exists such a semigroup, remains open.

\medskip
\textbf{5. Structure of the paper.}
Section~\ref{R} provides the representation theorem for the distortion semigroups.
Section~\ref{P} discusses how some mathematically and economically desirable properties of a
semigroup are expressed via its generator.
Section~\ref{L} deals with the problem of finding a semigroup
passing through a given concave distortion.
In Section~\ref{O}, we discuss some operations on semigroups.
Section~\ref{C} concludes.

\section{Representation of Distortion Semigroups}
\label{R}

In the theorem below we exclude the trivial cases
$\Psi_t(x)=x$, $t\ge0$, $x\in[0,1]$ and $\Psi_t(x)=1$, $t\ge0$, $x\in(0,1]$.

\begin{theorem}
\label{R1}
A family $(\Psi_t)_{t\ge0}$ is a concave distortion semigroup if
and only if there exists a concave function $G:(0,1)\to(0,\infty)$ such that
\begin{equation}
\label{r1}
\Psi_t(x)=\inf\Bigl\{y\in[x,1]:\int_x^y\frac{1}{G(s)}\,ds=t\Bigr\},
\quad t\ge0,\;x\in(0,1],
\end{equation}
where $\inf\emptyset=1$.
Moreover, the function $G$ is uniquely determined by the family
$(\Psi_t)_{t\ge0}$ via the relation
\begin{equation}
\label{r2}
G(x)=\lim_{t\downarrow0}\frac{\Psi_t(x)-x}{t},\quad x\in(0,1).
\end{equation}
\end{theorem}

\textit{Proof.}
Let us prove the ``if'' part.
It is obvious that each $\Psi_t$ is increasing, continuous, and $\Psi_t(1)=1$.
The semigroup property $\Psi_s\circ\Psi_t=\Psi_{s+t}$ is clear.

Fix $t>0$ and let us check that $\Psi_t$ is concave.
We can assume that $G$ is strictly concave (otherwise, we approximate
$G$ by strictly concave functions).
Suppose that $\Psi_t$ is not concave.
Denote $a=\sup\{x\in[0,1]:\Psi_t(x)<1\}$.
Then there exist $0<x_1<x_2<a$ and $\lambda\in(0,1)$ such that
$$
\Psi_t(\lambda x_1+(1-\lambda)x_2)
<\lambda\Psi_t(x_1)+(1-\lambda)\Psi_t(x_2).
$$
The relation
$$
\lim_{s\downarrow0}\frac{\Psi_s(x)-x}{s}=G(x),
$$
together with the strict concavity of~$G$, guarantees that
$$
\Psi_s(\lambda x_1+(1-\lambda)x_2)>
\lambda\Psi_s(x_1)+(1-\lambda)\Psi_s(x_2)
$$
for all sufficiently small~$s$.
This shows that the value
$$
t_0=\inf\{s\ge0:\Psi_s(\lambda x_1+(1-\lambda)x_2)\le
\lambda\Psi_s(x_1)+(1-\lambda)\Psi_s(x_2)\}
$$
satisfies $0<t_0<t$.
Due to the continuity of $\Psi_s(x)$ in~$s$, we have
$$
\Psi_{t_0}(\lambda x_1+(1-\lambda)x_2)
=\lambda\Psi_{t_0}+(1-\lambda)\Psi_{t_0}(x_2).
$$
The relations $x_1<x_2<a$ imply that $\Psi_t(x_1)<\Psi_t(x_2)<1$,
which, in turn, implies that $\Psi_{t_0}(x_1)<\Psi_{t_0}(x_2)<1$.
Then it is clear that
\begin{align*}
&\partial_s|_{s=t_0}[\Psi_s(\lambda x_1+(1-\lambda)x_2)
-\lambda\Psi_s(x_1)-(1-\lambda)\Psi_s(x_2)]\\[2mm]
&=G(\Psi_{t_0}(\lambda x_1+(1-\lambda)x_2))
-\lambda G(\Psi_{t_0}(x_1))-(1-\lambda)G(\Psi_{t_0}(x_2))\\[2mm]
&=G(\lambda\Psi_{t_0}(x_1)+(1-\lambda)\Psi_{t_0}(x_2))
-\lambda G(\Psi_{t_0}(x_1))-(1-\lambda)G(\Psi_{t_0}(x_2))>0,
\end{align*}
where the inequality follows from the strict concavity of~$G$.
But then for sufficiently small $\varepsilon>0$ we have
$$
\Psi_{t_0-\varepsilon}(\lambda x_1+(1-\lambda)x_2)
<\lambda\Psi_{t_0-\varepsilon}+(1-\lambda)\Psi_{t_0-\varepsilon}(x_2),
$$
which contradicts the choice of~$t_0$.

Let us prove the ``only if'' part.
First, we check that $\lim_{t\downarrow0}\Psi_t(x)=x$ for any $x\in[0,1]$.
Suppose that there exists $x_0$, for which this is wrong.
Obviously, $\Psi_t(x)$ is increasing in~$t$, so that there exists $a>x_0$ such
that $\Psi_t(x_0)>a$ for any $t>0$.
Due to the concavity of~$\psi_t$, we see that $\Psi_t\ge\Psi$ for any $t>0$,
where $\Psi$ is the piecewise linear function with $\Psi(0)=0$, $\Psi(x_0)=a$,
and $\Psi(1)=1$.
Then $\Psi_t=\Psi_{t/n}^n\ge\Psi^n$ for any $n\in\mathbb{N}$, where
$\Psi^n=\Psi\circ\dots\circ\Psi$ is the $n$-th composition of $\Psi$ with
itself.
It is clear that $\Psi^n(x)\to1$ for any $x\in(0,1]$, which brings us to a
contradiction.

Let us now prove that, for any $x\in(0,1)$, there exists
$\lim_{t\downarrow0}t^{-1}[\Psi_t(x)-x]$.
Fix $x\in(0,1)$ and $\varepsilon>0$.
It follows from the concavity of $\Psi_t$ that
$$
\frac{1-y}{1-x}\,(\Psi_t(y)-y)
\le\Psi_t(y)-y
\le\frac{y}{x}\,(\Psi_t(x)-x),\quad y\in[x,1],\;t\ge0.
$$
Combining this with the property $\lim_{t\downarrow0}\Psi_t(x)=x$, we see that
there exists $\delta>0$ such that
$$
(1-\varepsilon)(\Psi_t(x)-x)
\le\Psi_t(y)-y
\le(1+\varepsilon)(\Psi_t(x)-x),\quad
y\in[x,\Psi_t(x)],\;t\in(0,\delta).
$$
Then, for any $n\in\mathbb{N}$,
$$
(1-\varepsilon)\bigl(\Psi_{\frac{t}{n}}(x)-x\bigr)
\le\Psi_{\frac{t}{n}}(y)-y
\le(1+\varepsilon)\bigl(\Psi_{\frac{t}{n}}(x)-x\bigr),\quad
y\in[x,\Psi_t(x)],\;t\in(0,\delta).
$$
Consequently, for any $m\le n\in\mathbb{N}$,
$$
(1-\varepsilon)\bigl(\Psi_{\frac{t}{n}}(x)-x\bigr)
\le\Psi_{\frac{m}{n}t}(x)-\Psi_{\frac{m-1}{n}t}(x)
\le(1+\varepsilon)\bigl(\Psi_{\frac{t}{n}}(x)-x\bigr),\quad t\in(0,\delta).
$$
Summing up these inequalities, we get
$$
(1+\varepsilon)^{-1}\frac{\Psi_t(x)-x}{t}
\le\frac{\Psi_{t/n}(x)-x}{t/n}
\le(1-\varepsilon)^{-1}\frac{\Psi_t(x)-x}{t},\quad t\in(0,\delta),
$$
which implies that
\begin{align*}
&(1+\varepsilon)^{-1}\frac{\Psi_t(x)-x}{t}
\le\liminf_{n\to\infty}\frac{\Psi_{t/n}(x)-x}{t/n}\\
&\le\limsup_{n\to\infty}\frac{\Psi_{t/n}(x)-x}{t/n}
\le(1-\varepsilon)^{-1}\frac{\Psi_t(x)-x}{t},\quad t\in(0,\delta).
\end{align*}
As the function $\Psi_t(x)$ is increasing in~$t$, we get
\begin{align*}
&(1+\varepsilon)^{-1}\frac{\Psi_t(x)-x}{t}
\le\liminf_{s\downarrow0}\frac{\Psi_s(x)-x}{s}\\
&\le\limsup_{s\downarrow0}\frac{\Psi_{s}(x)-x}{s}
\le(1-\varepsilon)^{-1}\frac{\Psi_t(x)-x}{t},\quad t\in(0,\delta).
\end{align*}
In particular, we see that the ratio $s^{-1}[\Psi_s(x)-x]$ is bounded
by some constant for all~$s$ from some neighborhood of zero.
As $\varepsilon$ is arbitrary, we get
$$
\liminf_{s\downarrow0}\frac{\Psi_s(x)-x}{s}
=\limsup_{s\downarrow0}\frac{\Psi_{s}(x)-x}{s}
=\lim_{s\downarrow0}\frac{\Psi_{s}(x)-x}{s}.
$$

The function
$$
G(x)=\lim_{t\downarrow0}\frac{\Psi_t(x)-x}{t},\quad x\in(0,1)
$$
is positive and concave.
It follows from the semigroup property of $(\Psi_t)_{t\ge0}$ that
$$
\lim_{\varepsilon\downarrow0}\frac{\Psi_{t+\varepsilon}(x)-\Psi_t(x)}{\varepsilon}
=G(\Psi_t(x)),\quad t\ge0,\;x\in(0,1).
$$
Furthermore, the convergence $s^{-1}[\Psi_s(x)-x]$ to~$G(x)$ is uniform on any
compact subinterval of $(0,1)$ due to the concavity of the functions
$s^{-1}[\Psi_s(x)-x]$.
Therefore,
$$
\lim_{\varepsilon\downarrow0}\frac{\Psi_{t-\varepsilon}(x)-\Psi_t(x)}{\varepsilon}
=\lim_{\varepsilon\downarrow0}\frac{\Psi_{t-\varepsilon}(x)-
\Psi_\varepsilon(\Psi_{t-\varepsilon}(x))}{\varepsilon}
=-G(\Psi_t(x))
$$
for any $t>0$ and any $x>0$ such that $\Psi_t(x)<1$.
Thus, $\partial_t\Psi_t(x)=G(\Psi_t(x))$ for any $t>0$ and
any $x>0$ such that $\Psi_t(x)<1$.
As the function $G(\Psi_t(x))$ is continuous in~$t$ for $x>0$ such that
$\Psi_t(x)<1$, we get
$$
\Psi_t(x)=x+\int_0^t G(\Psi_s(x))ds,\quad t>0,\;x>0,\;\Psi_t(x)<1.
$$
Therefore,
$$
\int_x^{\Psi_t(x)}\frac{1}{G(s)}\,ds
=\int_0^t\frac{1}{G(\Psi_s(x))}\,d\Psi_s(x)
=\int_0^t\frac{G(\Psi_s(x))}{G(\Psi_s(x))}\,ds
=t,\quad t>0,\;x>0,\;\Psi_t(x)<1.
$$
This proves~\eqref{r1} for $t>0$ and $x>0$ such that $\Psi_t(x)<1$.

The semigroup property obviously implies that $\Psi_0(x)=x$, so
that~\eqref{r1} is trivially satisfied for $t=0$.
As both sides of~\eqref{r1} are increasing in~$x$ and take values in $[0,1]$,
we see that this equality is satisfied for all $x\in(0,1]$.~\hfill$\Box$

\medskip
The next lemma provides another representation of $(\Psi_t)_{t\ge0}$ in terms
of~$G$. It will be employed in Section~\ref{O}.
We denote by~$I$ the identity function $I(x)=x$ and denote by~$F^n$ the $n$-th
composition of~$F$ with itself, i.e. $F\circ\dots\circ F$.

\begin{lemma}
\label{R2}
Let $(\Psi_t)_{t\ge0}$ be a concave distortion with a generator~$G$.
Let $(G_n)$ be a sequence of concave functions $(0,1)\to(0,\infty)$
converging to~$G$ pointwise. Then
\begin{equation}
\label{r3}
\lim_{n\to\infty}\biggl(I+\frac{t}{n}\,G_n\biggr)^n(x)=\Psi_t(x),
\quad t\ge0,\;x\in(0,1],
\end{equation}
where $G_n$ is extended to~$\mathbb{R}$ by letting $G_n=0$ outside $(0,1)$.
\end{lemma}

\textit{Proof.}
Without loss of generality, $t=1$.
Fix $x>0$ such that $\Psi_1(x)<1$
and $\varepsilon>0$ such that $\Psi_1(x)+\varepsilon<1$.
We can find $\delta>0$, for which
\begin{equation}
\label{r4}
\Psi_1(x)-\varepsilon
\le\Psi_{1-\delta}(x)
\le\Psi_{1+\delta}(x)
\le\Psi_1(x)+\varepsilon
<1.
\end{equation}
Due to the concavity of $G_n$, the convergence $G_n\to G$ is
uniform on the interval $[x,\Psi_{1+\delta}(x)]$.
The same is true for the convergence
$n\bigl(\Psi_{\frac{1-\delta}{n}}-I\bigr)\to(1-\delta)G$ and
$n\bigl(\Psi_{\frac{1+\delta}{n}}-I\bigr)\to(1+\delta)G$.
Thus, there exists $N\in\mathbb{N}$ such that
$$
n\bigl(\Psi_{\frac{1-\delta}{n}}(y)-y\bigr)
\le G_n(y)
\le n\bigl(\Psi_{\frac{1+\delta}{n}}(y)-y\bigr),\quad
y\in[x,\Psi_{1+\delta}(x)],\;n\ge N.
$$
This means that
$$
\Psi_{\frac{1-\delta}{n}}(y)
\le y+\frac{G_n(y)}{n}
\le\Psi_{\frac{1+\delta}{n}}(y),\quad
y\in[x,\Psi_{1+\delta}(x)],\;n\ge N.
$$
It is easy to see by the induction in~$k$ that
$$
x\le\Psi_{(1-\delta)\frac{k}{n}}(x)
\le\biggl(I+\frac{G_n}{n}\biggr)^k(x)
\le\Psi_{(1+\delta)\frac{k}{n}}(x)
\le\Psi_{1+\delta}(x)
\quad n\ge N,\;k=1,\dots,n.
$$
Combining this inequality for $k=n$ with~\eqref{r4},
we get~\eqref{r3} for $x>0$ such that $\Psi_1(x)<1$.

Furthermore, it is easy to see that both
$\limsup_n\bigl(I+\frac{G_n}{n}\bigr)^n$ and
$\liminf_n\bigl(I+\frac{G_n}{n}\bigr)^n$ are increasing functions
taking values in $[0,1]$.
As $\Psi_1(x)$ is also increasing and takes values in $[0,1]$, we
conclude that~\eqref{r3} is true for any $x\in(0,1]$.~\hfill$\Box$

\medskip
To conclude this section, let us consider four performance measures:
\begin{mitemize}
\item Sharpe ratio $\text{SR}(X)=\mathsf{E}X/\sigma(X)$,
where $\sigma$ is the standard deviation;
\item Risk-Adjusted Return on Capital $\text{RAROC}(X)=\mathsf{E}X/\text{V@R}(X)$,
where V@R is the Value at Risk;
\item Gain-Loss Ratio $\text{GLR}(X)=\mathsf{E}X/\mathsf{E}X^-$, where
$X^-=\max\{-X,0\}$;
\item Coherent Risk-Adjusted Return on Capital
$\text{CRAROC}(X)=\mathsf{E}X/\rho(X)$, where $\rho$ is a coherent risk measure.
\end{mitemize}
As shown in~\cite{CM071}, SR and RAROC are not coherent indices.
GLR is a coherent index, but it does not belong to the AIW class.
CRAROC is a coherent index. If $\rho$ belongs to the Weighted V@R class,
i.e.
$$
\rho(X)=-\int_\mathbb{R}xd(\Psi(D_X(x)))
$$
with some concave distortion~$\Psi$, then CRAROC is an AIW index
corresponding~to
$$
\Psi_t(x)=\frac{1}{1+t}\,x+\frac{t}{1+t}\,\Psi(x),\quad t\ge0,\;x\in[0,1].
$$
However, this is not a distortion semigroup.
Indeed, any distortion semigroup (except for the trivial one $\Psi_t(x)=x$) has
the property $\lim_{t\to\infty}\Psi_t(x)=1$ for $x\in(0,1]$; the family given
above does not have this property.

Thus, neither of the above performance measures is an AIW index corresponding
to a distortion semigroup. Nevertheless, nice representatives of such indices
do exist and some of them are given by~\eqref{i3}--\eqref{i6}.
The generators for these semigroups are, respectively:
\begin{align*}
G_1(x)&=x,\\
G_2(x)&=-(1-x)\ln(1-x),\\
G_3(x)&=-x\ln x,\\
G_4(x)&=\frac{1}{\sqrt{2\pi}}\,e^{-\frac{(\Phi^{-1}(x))^2}{2}}.
\end{align*}
The aim of the next section is to figure out which of the corresponding
indices have better properties.

\section{Properties of Distortion Semigroups}
\label{P}

\subsection{Mathematical Properties}

Let $(\Psi_t)_{t\ge0}$ be a semigroup with generator~$G$, and define $G(0)=G(1):=0$.
It is obvious that
\begin{equation}
\label{p1}
\Psi_t(0+)=0\;\Longleftrightarrow\;
\int_0^\varepsilon\frac{1}{G(s)}\,ds=\infty;
\end{equation}
we do not specify the values of $t>0$ and $\varepsilon\in(0,1)$ here since
clearly the above properties do not depend on the choice of~$t$ and
$\varepsilon$.
Furthermore,
\begin{equation}
\label{p2}
\Psi_t<1\text{ on }[0,1)\;\Longleftrightarrow\;
\int_{1-\varepsilon}^1\frac{1}{G(s)}\,ds=\infty.
\end{equation}
The theorem below establishes some finer properties of $(\Psi_t)_{t\ge0}$ in
terms of generators. By $f'_+$ ($\partial_x^+ f$) and $f'_-$ ($\partial_x^-f$) we denote the
right and the left (partial) derivatives (in~$x$) of a function $f$, respectively.

\begin{theorem}
\label{P1}
{\bf(i)} For any $t>0$, the function $\Psi_t$ is strictly concave on the set
${\{\Psi_t<1\}}$ if and only if $G$ is strictly concave.

{\bf(ii)} Assume that $\int_0^\varepsilon\frac{1}{G(s)}\,ds=\infty$.
Then $\partial_x^+\Psi_t(0)=e^{G'_+(0)t}$ for any $t\ge0$.

{\bf(iii)} Assume that $G(1-)=0$. Then $\partial_x^-\Psi_t(1)=e^{G'_-(1)t}$
for any $t\ge0$.
\end{theorem}

\textit{Proof.}
{\bf(i)} The ``if'' part was verified in the proof of Theorem~\ref{R1}.
Let us prove the ``only if'' part.
Suppose that $G$ is not strictly concave.
Then there exists an interval $[a,b]\subseteq(0,1)$ such that
$G(x)=\alpha+\beta x$ on $[a,b]$.
The direct application of~\eqref{r1} shows that, for $x\in[a,b]$ such that
$\Psi_t(x)$ also belongs to $[a,b]$, we have
$$
\Psi_t(x)=\beta^{-1}(\alpha+\beta x)e^{\beta t}-\beta^{-1}\alpha.
$$
In view of the continuity of $\Psi_t(x)$ in~$t$, there exists a sufficiently
small $t>0$ such that $\Psi_t$ is linear on some interval.
This is a contradiction.

{\bf(ii)} Consider first the case, where $G'_+(0)<\infty$.
Fix $\varepsilon>0$ and $t>0$.
The equality $\int_x^{e^{G'_+(0)t}x}\frac{1}{G'_+(0)s}\,ds=t$ guarantees that
$$
\int_x^{(1-\varepsilon)e^{G'_+(0)t}x}\frac{1}{G(s)}\,ds<t
<\int_x^{(1+\varepsilon)e^{G'_+(0)t}x}\frac{1}{G(s)}\,ds
$$
for all sufficiently small~$x$ (note that $G(0+)=0$). Then
$$
(1-\varepsilon)e^{G'_+(0)t}x<\Psi_t(x)<(1+\varepsilon)e^{G'_+(0)t}x,
$$
which implies that $\partial_x^+\Psi_t(0)=e^{G'_+(0)t}$.

Consider now the case $G'_+(0)=\infty$.
Take concave functions~$G_n$ that increase to~$G$.
Then automatically $G'_{n+}(0)\to\infty$.
The functions $\Psi_{t,n}$ defined via~\eqref{r1} with $G$ replaced by~$G_n$
increase to~$\Psi_t$.
The condition $\int_0^\varepsilon\frac{1}{G(s)}\,ds=\infty$ ensures that
$\Psi_t(0+)=0$, and hence, $\Psi_{t,n}(0+)=0$.
Thus, $\partial_x^+\Psi_t(0)\ge\partial_x^+\Psi_{t,n}(0)$ for any~$n$, and the
latter quantities tend to infinity as proved above.

{\bf(iii)} The case, where $\int_{1-\varepsilon}^1\frac{1}{G(s)}\,ds=\infty$,
is treated similarly as~(ii).
If $\int_{1-\varepsilon}^1\frac{1}{G(s)}\,ds<\infty$, then $G'_-(1)=-\infty$ and
$\Psi_t=1$ in some neighborhood of~1, so that
$\partial_x^-\Psi_t(1)=0=e^{G'_-(1)t}$.~\hfill$\Box$

\subsection{Economical Properties}

Consider a coherent index~$\alpha$ defined on the space $L^0$ of all random
variables via~\eqref{i2}, where $\int_\mathbb{R}xd(\Psi_t(D_X(x)))$ is understood as
$\int_\mathbb{R}x^+d(\Psi_t(D_X(x)))-\int_\mathbb{R}x^-d(\Psi_t(D_X(x)))$ with the
convention $\infty-\infty=-\infty$ (recall that $x^+=\max\{x,0\}$ and
$x^-=\max\{-x,0\}$).
This way of extending indices to $L^0$ is proposed by the corresponding
extension of coherent risks to~$L^0$, which turns out to be rather convenient,
in particular, in application to pricing; see~\cite{C071}.
It follows from~\cite[Th.~3.3, 46]{C06} that
\begin{equation}
\label{p3}
\int_\mathbb{R}xd(\Psi_t(D_X(x)))
=\inf_{\mathsf{Q}\in\mathcal{D}_t}\mathsf{E}_\mathsf{Q}X,\quad X\in L^0,
\end{equation}
where the expectation $\mathsf{E}_\mathsf{Q}X$ is understood as
$\mathsf{E}_\mathsf{Q}X^+-\mathsf{E}_\mathsf{Q}X^-$ (again with the convention
$\infty-\infty=-\infty$) and
\begin{align}
\label{p4}
\mathcal{D}_t&=\{\mathsf{Q}:\mathsf{E}(d\mathsf{Q}/d\mathsf{P}-x)^+\le\Phi_t(x)\;
\forall x\in\mathbb{R}_+\},\\
\label{p5}
\Phi_t(x)&=\sup_{y\in[0,1]}[\Psi_t(y)-xy],\quad x\in\mathbb{R}_+.
\end{align}
This provides the representation of~$\alpha$ in the form~\eqref{i1}.

In what follows, we will consider the space
$$
L^1(\alpha)=\Bigl\{X\in L^0:\int_\mathbb{R}xd(\Psi_t(D_X(x)))>-\infty\;\;
\text{and }\int_\mathbb{R}xd(\Psi_t(D_{-X}(x)))>-\infty\;\forall t\ge0\Bigr\}.
$$
It is easy to see from~\eqref{p3} that this is a linear space. If
$\partial_x^+\Psi_t(0)<\infty$ for any~$t\ge0$ (as is the case for
examples~\eqref{i3}, \eqref{i4}), then $L^1(\alpha)=L^1$. In
general, we always have the inclusions $L^\infty\subseteq
L^1(\alpha)\subseteq L^1$.

\bigskip
\textbf{1. Basic properties.}
First of all, let us consider four economically desirable properties of
coherent indices discussed in~\cite{CM071} (in the conditions below,
$X,Y\in L^1(\alpha)$):
\begin{mitemize}
\item (Law invariance) If $X$ and $Y$ have the same law, then
$\alpha(X)=\alpha(Y)$;
\item (Consistency with second-order stochastic dominance)
If $Y$ second order stochastically dominates~$X$
(i.e. $\mathsf{E}U(X)\le\mathsf{E}U(Y)$ for all increasing
concave functions~$U$), then $\alpha(X)\le\alpha(Y)$;
\item (Arbitrage consistency) $X\ge0$ if and only if $\alpha(X)=\infty$;
\item (Expectation consistency) if $\mathsf{E}X<0$, then $\alpha(X)=0$;
if $\mathsf{E}X>0$, then $\alpha(X)>0$.
\end{mitemize}
Obviously, any AIW index is law invariant.
As shown in~\cite[Sect.~3]{CM071}, it is also second-order monotone.
Arbitrage consistency for an AIW index defined by a family $(\Psi_t)_{t\ge0}$
is equivalent to the property $\lim_{t\to\infty}\Psi_t(x)=1$ for $x\in(0,1]$;
this is satisfied by any distortion semigroup, except for the trivial one
$\Psi_t(x)=x$.
Finally, the expectation consistency for an AIW index is equivalent to the
property $\Psi_0(x)=x$, which is satisfied by any distortion semigroup, except
for the trivial one $\Psi_t(x)=1$.

To sum up, the above four properties are automatically satisfied by
distortion semigroups.
Below we investigate finer economical properties, which are satisfied only by
some semigroups.
We fix a semigroup $(\Psi_t)_{t\ge0}$ with a generator~$G$
and define~$\alpha$ via~\eqref{i2}.

\bigskip
\textbf{2. Positivity on unbounded random variables.}
The proposition below shows that the condition
$\int_0^\varepsilon\frac{1}{G(s)}\,ds=\infty$ is extremely desirable.
Namely, if it is violated, then $\alpha(X)=0$ for any~$X$ unbounded below,
which is an unacceptable property because most of the distributions
employed in financial modelling (e.g. the Gaussian one) have unbounded support.

\begin{proposition}
\label{P2}
Suppose that the probability space is atomless.
The following conditions are equivalent:
\begin{mitemize}
\item[\rm(a)] there exists $X$ with $\mathop{\rm essinf}_\omega X(\omega)=-\infty$
such that $\alpha(X)>0${\rm;}
\item[\rm(b)] for any $X\in L^1(\alpha)$ and any $t\ge0$,
$\mathop{\rm argmin}_{\mathsf{Q}\in\mathcal{D}_t}\mathsf{E}_\mathsf{Q}X\ne\emptyset${\rm;}
\item[\rm(c)] for any $t\ge0$, $\Psi_t(0+)=0${\rm;}
\item[\rm(d)] $\int_0^\varepsilon\frac{1}{G(s)}\,ds=\infty$.
\end{mitemize}
\end{proposition}

\textit{Proof.}
(a)$\Rightarrow$(c)
If $\Psi_t(0+)>0$ for some $t$, then the same is true for any~$t$.
For any $X$ unbounded below, we then have
$\int_\mathbb{R}x^-d(\Psi_t(D_X(x)))=\infty$, $t>0$.
According to our agreement, $\int_\mathbb{R}xd(\Psi_t(D_X(x)))=-\infty$, $t>0$,
which means that $\alpha(X)=0$.

(c)$\Rightarrow$(a)
Fix $t>0$.
Clearly, we can find $X$ unbounded below and such that
$\int_\mathbb{R}xd(\Psi_t(D_X(x)))\ge0$.
Then $\alpha(X)\ge t$.

(c)$\Rightarrow$(b) If $\Psi_t(0+)=0$,
then the function~$\Phi_t$ given by~\eqref{p5} satisfies $\Phi_t(\infty)=0$,
which means that the set~$\mathcal{D}_t$ given by~\eqref{p4} is uniformly
integrable. Now, the desired statement follows from~\cite[Prop.~2.9]{C071}.

(b)$\Rightarrow$(c) Suppose that $\Psi_t(0+)>0$.
It is easy to see that then
$\mathcal{D}_t=L^0+(1-\Psi_t(0+))\widetilde{\mathcal{D}}_t$ (we identify here
probability measures with their Radon-Nikodym derivatives with respect
to~$\mathsf{P}$), where $\widetilde{\mathcal{D}}_t$ is given
by~\eqref{p4}--\eqref{p5} with $\Psi_t$ replaced by
$\widetilde\Psi_t=(1-\Psi_t(0+))^{-1}\Psi_t$.
Let $X$ be uniformly distributed on $[0,1]$.
The previous implication guarantees that
$\mathop{\rm argmin}_{\mathsf{Q}\in\widetilde{\mathcal{D}}_t}
\mathsf{E}_\mathsf{Q}X\ne\emptyset$.
However, $\mathop{\rm argmin}_{Z\in L^0}\mathsf{E}ZX=\emptyset$.
As a result, $\mathop{\rm argmin}_{\mathsf{Q}\in\mathcal{D}_t}
\mathsf{E}_\mathsf{Q}X=\emptyset$.

(c)$\Leftrightarrow$(d) This is equivalence~\eqref{p1}.~\hfill$\Box$

\bigskip
\textbf{3. Strict quasi-concavity.}
The next proposition shows that the strict concavity of~$G$, together with the
condition $\int_{1-\varepsilon}^1\frac{1}{G(s)}\,ds=\infty$, is responsible
for the following property of~$\alpha$, which might be termed the strict
quasi-concavity: if $\alpha(X)=\alpha(Y)$ and $X,Y$ are not comonotone, then
$\alpha(X+Y)>\alpha(X)$.
Recall that random variables $X,Y$ are called \textit{comonotone} if there
exists a random variable~$Z$ and increasing functions $f,g$ such that
$X=f(Z)$, $Y=g(Z)$.

\begin{proposition}
\label{P3}
Suppose that the probability space is atomless.
The following conditions are equivalent:
\begin{mitemize}
\item[\rm(a)] for any $X,Y\in L^1(\alpha)$ that are not comonotone and are
such that $0<\alpha(X)=\alpha(Y)<\infty$, we have $\alpha(X+Y)>\alpha(X)$.
\item[\rm(b)] for any $t>0$, $\Psi_t$ is strictly concave;
\item[\rm(c)] $G$ is strictly concave and
$\int_{1-\varepsilon}^1\frac{1}{G(s)}\,ds=\infty$.
\end{mitemize}
\end{proposition}

\textit{Proof.}
(a)$\Rightarrow$(b)
Suppose that, for some $t>0$, $\Psi_t$ is not strictly concave.
Then there exists an interval $[a,b]$, on which $\Psi_t$ is affine.
As the probability space is atomless, it supports a random variable~$X$ with a
uniform distribution on $[0,1]$.
The random variable
$$
Y=\begin{cases}
X&\text{on}\;\;\{X<a\}\cup\{X>b\},\\
a+b-X&\text{on}\;\;\{a\le X\le b\}
\end{cases}
$$
also has the uniform distribution, so that
$$
\int_\mathbb{R}xd(\Psi_t(D_X(x)))
=\int_\mathbb{R}xd(\Psi_t(D_Y(x)))
=\int_0^1 xd\Psi_t(x).
$$
Then
\begin{align*}
\int_0^1 xd\bigl(\Psi_t\bigl(D_{\frac{X+Y}{2}}(x)\bigr)\bigr)
&=\int_0^a xd\Psi_t(x)
+\frac{a+b}{2}(\Psi_t(b)-\Psi_t(a))
+\int_b^1 xd\Psi_t(x)\\
&=\int_0^a xd\Psi_t(x)
+\int_a^b\frac{a+b}{2}d\Psi_t(x)
+\int_b^1 xd\Psi_t(x)\\
&=\int_0^1 xd\Psi_t(x),
\end{align*}
where the third equality follows from the affinity of~$\Psi_t$ on $[a,b]$.
Consider now the random variables $\widetilde X=X-\int_0^1 xd\Psi_t(x)$ and
$\widetilde Y=Y-\int_0^1 xd\Psi_t(x)$.
As $\Psi_t$ is strictly increasing in~$t$, we have
$$
\int_\mathbb{R}xd(\Psi_{t+\varepsilon}(D_{\widetilde X}(x)))<0,\qquad
\int_\mathbb{R}xd(\Psi_{t+\varepsilon}(D_{\widetilde Y}(x)))<0
$$
for any $\varepsilon>0$, so that $\alpha(\widetilde X)=\alpha(\widetilde Y)=t$.
Furthermore,
$$
\int_\mathbb{R}xd\bigl(\Psi_t\bigl(D_{\frac{\widetilde X+\widetilde Y}{2}}(x)\bigr)\bigr)
=\int_\mathbb{R}xd\bigl(\Psi_t\bigl(D_{\frac{X+Y}{2}}(x)\bigr)\bigr)
-\int_0^1 xd\Psi_t(x)=0,
$$
and hence,
$\alpha(\widetilde X+\widetilde Y)
=\alpha\bigl(\frac{\widetilde X+\widetilde Y}{2}\bigr)=t$.
Clearly, $\widetilde X$ and $\widetilde Y$ are not comonotone, so we have got
a contradiction with~(a).

(b)$\Rightarrow$(a)
Let $t_0=\alpha(X)=\alpha(Y)$.
Integration by parts yields the representation
$$
\int_\mathbb{R}xd(\Psi_t(D_X(x)))
=-\int_{-\infty}^0\Psi_t(D_X(x))dx+\int_0^\infty(1-\Psi_t(D_X(x)))dx.
$$
If $X$ is bounded, we immediately see that this
function is continuous in~$t$. This means the continuity of the function
$t\mapsto\inf_{\mathsf{Q}\in\mathcal{D}_t}\mathsf{E}_\mathsf{Q}X$.
Approximating now $X\in L^1(\alpha)$ by its truncations, we see that the same
continuity property holds true for such~$X$.
Hence,
$$
\int_\mathbb{R}xd(\Psi_{t_0}(D_X(x)))
=\int_\mathbb{R}xd(\Psi_{t_0}(D_Y(x)))
=0.
$$
It follows from~\cite[Th.~5.1]{C06} that
$\int_\mathbb{R}xd(\Psi_{t_0}(D_{X+Y}(x)))>0$.
Employing the continuity of the map
$t\mapsto\int_\mathbb{R}xd(\Psi_t(D_{X+Y}(x)))$, we get that
$\alpha(X+Y)>t_0$.

(b)$\Leftrightarrow$(c)
This equivalence follows from Theorem~\ref{P1}~(i) combined
with~\eqref{p2}.~\hfill$\Box$

\medskip
As an application of the previous proposition, consider the following
optimization problem. Let $S_0=(S_0^1,\dots,S_0^d)\in\mathbb{R}^d$ denote the initial prices
of $d> 1$ assets and $S_1=(S_1^1,\dots,S_1^d)\in (L^1(\alpha))^d$ their terminal
discounted prices. Consider the portfolio optimization problem
\begin{equation}
\label{p6}
\alpha(\langle h,S_1-S_0\rangle)\longrightarrow\max,
\quad h\in\mathbb{R}^d\setminus\{0\}.
\end{equation}
Recall that the model is arbitrage-free if there exists no $h\in\mathbb{R}^d$
such that $\mathsf{P}(\langle h,S_1-S_0\rangle\ge0)=1$ and
$\mathsf{P}(\langle h,S_1-S_0\rangle>0)>0$.

\begin{corollary}
\label{P4}
Suppose that $\mathsf{E}S_1\ne S_0$ and there is no arbitrage.
Assume that $S_1^1,\dots,S_1^d$ have a joint density.
Let $(\Psi_t)_{t\ge0}$ satisfy the equivalent conditions of
Proposition~\ref{P3}.
Then the optimal value in~\eqref{p6} is strictly positive and finite,
there exists a solution of~\eqref{p6},
and it is unique up to multiplication by a positive constant.
\end{corollary}

\textit{Proof.}
Denote by $a_*$ the optimal value in~\eqref{p6}.
Clearly, it remains the same if the optimization is performed over the unit
sphere~$S$.
Choose $h_n\in S$ with $\alpha(\langle h_n,S_1-S_0\rangle)\to a_*$.
Passing on to a subsequence, we can assume that $h_n$ converge to $h_*\in S$.
As explained in the preceding proof the map
$t\mapsto\int_\mathbb{R}xd(\Psi_t(D_X(x)))$ is continuous and is strictly
decreasing for any non-degenerate $X\in L^1(\alpha)$.
Clearly,
$$
\int_\mathbb{R}xd(\Psi_t(D_{X_n}(x)))\xrightarrow[n\to\infty]{}
\int_\mathbb{R}xd(\Psi_t(D_{X_*}(x))),\quad t\ge0,
$$
where $X_n=\langle h_n,S_1-S_0\rangle$, $X_*=\langle h_*,S_1-S_0\rangle$.
Hence, $\alpha(X_n)\to\alpha(X_*)$, so that $X_*$ is optimal for~\eqref{p6}.

As $S_1^1,\dots,S_1^d$ has a joint density, $X_*$ is non-degenerate.
Due to the absence of arbitrage, $\mathsf{P}(X_*<0)>0$.
Obviously, $\Psi_t(x)\xrightarrow[t\to\infty]{}1$ for any $x\in(0,1]$.
Hence,
$$
\lim_{t\to\infty}\int_\mathbb{R}xd(\Psi_t(D_{X_*}(x)))<0,
$$
which implies that $\alpha(X_*)<\infty$.
As $\int_\mathbb{R}xd(\Psi_0(D_X(x)))=\mathsf{E}X$, we see that
$\alpha(X)=0$ for any $X\in L^1(\alpha)$ with $\mathsf{E}X\le0$;
$\alpha(X)>0$ for any $X\in L^1(\alpha)$ with $\mathsf{E}X>0$.
In view of the condition $\mathsf{E}(S_1-S_0)\ne0$, we get $\alpha(X_*)>0$.

Suppose now that there exists $h'\in S$ that is not a positive multiple
of~$h_*$ such that $\alpha(X')=a_*$, where $X'=\langle h',S_1-S_0\rangle$.
Then the vector $(X_*,X')$ has a joint density, so that $X_*,X'$ are not
comonotone.
According to Proposition~\ref{P3}, $\alpha(X_*+X')>a_*$, which is a
contradiction.~\hfill$\Box$

\bigskip
\textbf{4. Left tail of extreme measure densities.}
Let $(\Psi_t)_{t\ge0}$ be a semigroup satisfying the equivalent
conditions of Proposition~\ref{P2}.
As shown by that proposition, for any ${X\in L^1(\alpha)}$ and any $t\ge0$, the
set of \textit{extreme measures}
$\mathop{\rm argmin}_{\mathsf{Q}\in\mathcal{D}_t}\mathsf{E}_\mathsf{Q}X$
is non-empty. If $X$ has a continuous distribution, then this set consists of
a unique measure $\mathsf{Q}_t^*(X)$ given by
$d\mathsf{Q}_t^*(X)/d\mathsf{P}=(\partial_x\Psi_t)(D_X(X))$
(see~\cite[Sect.~6]{C06}; it is not important whether we take the left or right
partial derivative here as the random variable $D_X(X)$ is uniformly
distributed).
This measure has certain economical similarities (see~\cite[Sect.~2]{CM071})
with the state-price density based on the classical expected utility,
i.e. with the measure~$\mathsf{Q}$ given by $d\mathsf{Q}/d\mathsf{P}=cU'(W)$,
where ${U:\mathbb{R}\to\mathbb{R}}$ is a concave increasing function,
$W$ is the terminal wealth of a position, and $c$ is the normalizing constant.
A desirable property of the classical utility~$U$ is that
${U'(-\infty)=\infty}$, which means that $cU'(W)$ is unbounded (we assume here
that $W$ is unbounded below).
By analogy, a desirable property of extreme measures would be
the unboundedness of $(\partial_x\Psi_t)(D_X(X))$, which, in turn,
is equivalent to the property $\partial_x^+\Psi_t(0)=\infty$.

The unboundedness of $(\partial_x\Psi_t)(D_X(X))$ means that large losses are exaggerated
up to infinity. As argued above, this is an economically desirable feature.
However, it bears the potential danger that in applications the value of the index
might depend essentially on the tail of the distribution, which, in turn, is
unstable under changes of the data set, a change of the model, etc. But let us
remark in this connection that Eberlein and Madan~\cite{EM07} performed some
data tests for four coherent indices from~\cite{CM071} for the performance of
hedge funds. The four indices were AIMIN, AIMAX, AIMAXMIN, and AIMINMAX.
For the first of these indices, the extreme measure densities
$(\partial_x\Psi_t)(D_X(X))$ are bounded; for the other three indices, these
densities are unbounded. The tests showed high correlation between the results for
the last three indices, while the results for AIMIN were relatively
uncorrelated with those three indices. This suggests the superiority of
coherent indices with unbounded densities of extreme measures.

The following statement is a direct consequence of Theorem~\ref{P1}.

\begin{corollary}
\label{P5}
Suppose that the probability space is atomless and the equivalent conditions
of Proposition~\ref{P2} are satisfied.
The following conditions are equivalent:
\begin{mitemize}
\item[\rm(a)] for any $X\in L^1(\alpha)$ with a continuous distribution and
any $t>0$, the density $d\mathsf{Q}_t^*(X)/d\mathsf{P}$ is unbounded;
\item[\rm(b)] for any $t>0$, $\partial_x^+\Psi_t(0)=\infty${\rm;}
\item[\rm(c)] $G'_+(0)=\infty$.
\end{mitemize}
\end{corollary}

\smallskip
\textbf{5. Right tail of extreme measure densities.} Most well-known
utility functions (e.g. CARA, HARA, etc.) satisfy the property
$U'(\infty)=0$. This suggests that it is desirable that the
classical state-price densities $cU'(W)$ should have its essential
infimum equal to zero, and the same would be desirable for the
extreme measure densities.

The following statement is a direct consequence of Theorem~\ref{P1}.

\enlargethispage{\baselineskip}
\begin{corollary}
\label{P6}
Suppose that the probability space is atomless and the equivalent conditions
of Proposition~\ref{P2} are satisfied.
The following conditions are equivalent:
\begin{mitemize}
\item[\rm(a)] for any $X\in L^1(\alpha)$ with a continuous distribution and
any $t>0$, the density $d\mathsf{Q}_t^*(X)/d\mathsf{P}$ has
essential infimum equal to zero;
\item[\rm(b)] for any $t>0$, $\partial_x^-\Psi_t(1)=0${\rm;}
\item[\rm(c)] $\int_{1-\varepsilon}^1\frac{1}{G(s)}\,ds<\infty$ or $G'_-(1)=-\infty$.
\end{mitemize}
\end{corollary}

\textbf{6. Examples.}
We have considered four economically meaningful properties of distortion
semigroups:
\begin{mitemize}
\item[\bf I.] for any $t>0$, $\Psi_t(0+)=0$;
\item[\bf II.] for any $t>0$, $\Psi_t$ is strictly concave;
\item[\bf III.] for any $t>0$, $\partial_x^+\Psi_t(0)=\infty$;
\item[\bf IV.] for any $t>0$, $\partial_x^-\Psi_t(1)=0$.
\end{mitemize}
The first property is extremely important both from the mathematical
and from the economical perspectives. The second one is desirable
mathematically. The third and fourth properties are economically
reasonable.

The table below shows which of those properties are satisfied for
semigroups~\eqref{i3}--\eqref{i6}.
The table suggests that the best properties are exhibited by
semigroups~\eqref{i5} and~\eqref{i6}.

\begin{center}
\begin{tabular}{|c||c|c|c|c|}
\hline
\rule{0mm}{5mm}&\bf I&\bf II&\bf III&\bf IV\\
\hhline{|=||=|=|=|=|}
\rule{0mm}{5mm}\eqref{i3}&$+$&$-$&$-$&$-$\\[1mm]
\hline
\rule{0mm}{5mm}\eqref{i4}&$+$&$+$&$-$&$-$\\[1mm]
\hline
\rule{0mm}{5mm}\eqref{i5}&$+$&$+$&$+$&$-$\\[1mm]
\hline
\rule{0mm}{5mm}\eqref{i6}&$+$&$+$&$+$&$+$\\[1mm]
\hline
\end{tabular}

\nopagebreak
\vspace{5mm}
{\small\textbf{Table~1.} Properties I--IV for
semigroups~\eqref{i3}--\eqref{i6}}
\end{center}

\section{Logarithm of a Concave Distortion}
\label{L}

Let $\Psi$ be a concave distortion.
In this section we will consider the problem of the existence and the
uniqueness of a distortion semigroup $(\Psi_t)_{t\ge0}$ such that $\Psi_1=\Psi$.
The generator of such a semigroup might be called the \textit{logarithm} of~$\Psi$.

The problem of recovering the generator by knowing the whole semigroup
is trivial as the generator is given by~\eqref{r2}.
In contrast, the problem of recovering the generator by knowing only $\Psi_1$
is not trivial, and we have only very partial results in this
direction.\footnote{There exists a slight analogy between our situation and
the problem of recovering the diffusion coefficient $\sigma(x)$ of a
one-dimensional diffusion $dX_t=\sigma(X_t)dB_t$, $X_0=0$ by the marginal
distributions of~$X$. If for any $t>0$ we are given the density $p_t(x)$ of~$X_t$,
then there exists a simple formula for~$\sigma$; see Dupire~\cite{D94}.
However, if we are given only the density~$p_1(x)$, then the problem of the
existence and the uniqueness of~$\sigma$ is very hard and is open, to the best
of our knowledge.
Let us mention in this connection the paper by Israel, Rosenthal, and
Wei~\cite{IRW01}, where the generator of a Markov chain with discrete state
space is found by the knowledge of its time-1 transition matrix.}
The next theorem states that there exists at most one
distortion semigroup $(\Psi_t)_{t\ge0}$ with $\Psi_1=\Psi$, under the additional assumption
$\Psi'_-(1)>0$. This condition is linked to the behavior of the right tail of
extreme measure densities as discussed in the previous section.

\begin{theorem}
\label{L1}
Let $\Psi$ be a concave distortion such that $\Psi'_-(1)>0$.
Then there exists at most one concave distortion semigroup $(\Psi_t)_{t\ge0}$
with $\Psi_1=\Psi$. If it exists, then its generator is given~by
\begin{equation}
\label{l1}
G(x)=\lim_{n\to\infty}\frac{\ln\Psi_-'(1)(\Psi^n(x)-1)}%
{\Psi'(x)\dots\Psi'(\Psi^{n-1}(x))},\quad x\in(0,1),
\end{equation}
where $\Psi^n=\Psi\circ\dots\circ\Psi$ is the $n$-th composition of~$\Psi$
with itself.
\end{theorem}

\textit{Proof.}
Let $(\Psi_t)_{t\ge0}$ be a distortion semigroup with $\Psi_1=\Psi$.
Let $G$ be its generator.
The condition $\Psi'_-(1)>0$ ensures that $\Psi<1$ on $[0,1)$, which means that
$\int_{1-\varepsilon}^1\frac{1}{G(s)}\,ds=\infty$.
Hence, $\int_x^{\Psi(x)}\frac{1}{G(s)}\,ds=1$ for any $x\in(0,1)$.
Differentiating both sides in~$x$, we get
$$
\frac{1}{G(x)}=\frac{\Psi'(x)}{G(\Psi(x))},\quad x\in(0,1).
$$
Iterating this equality, we get
$$
G(x)=\frac{G(\Psi(x))}{\Psi'(x)}
=\frac{G(\Psi^2(x))}{\Psi'(x)\Psi'(\Psi(x))}
=\dots=\frac{G(\Psi^n(x))}{\Psi'(x)\dots\Psi'(\Psi^{n-1}(x))},
\quad x\in(0,1),\;n\in\mathbb{N}.
$$
Theorem~\ref{P1}~(iii) implies that $G'_-(1)$ is finite and is equal to
$\ln\Psi'_-(1)$. Clearly, ${\Psi^n(x)\to1}$ and $\Psi^n(x)<1$ for $x\in(0,1)$.
Therefore,
$$
\lim_{n\to\infty}\frac{G(\Psi^n(x))}{\Psi^n(x)-1}
=G'_-(1)=\ln\Psi'_-(1),\quad x\in(0,1).
$$
This proves~\eqref{l1}, from which the uniqueness of~$G$, and thus, the
uniqueness of $(\Psi_t)_{t\ge0}$ is obvious.~\hfill$\Box$

\medskip
The next example shows that a distortion semigroup $(\Psi_t)_{t\ge0}$ with
$\Psi_1=\Psi$ might not exist in some cases.

\begin{example}\rm
\label{L2}
Take $0<a<b<1$ and consider a concave piecewise linear function~$\Psi$ with
$\Psi(0+)=0$, $\Psi(a)=b$, $\Psi(1)=1$ (note that $\Psi'_-(1)>0$, so that it
satisfies the condition of the previous theorem).
Then $\Psi'$ equals~$\frac{b}{a}$ on $(0,a)$ and is equal to
$\frac{1-b}{1-a}$ on $(a,1)$. Thus, $\Psi'(x)$ suffers a decrease in
$\frac{(1-a)b}{(1-b)a}>1$ times at the point~$a$.
For any $n\ge2$, $\Psi^n(a)>a$, so that $\Psi'(\Psi^n(x))$ is continuous at
the point~$a$. The function $\Psi^n(x)-1$ is continuous everywhere.
As a result, for any $n\in\mathbb{N}$, the function
\begin{equation}
\label{l2}
G_n(x)=\frac{\ln\Psi'_-(1)(\Psi^n(x)-1)}%
{\Psi'(x)\dots\Psi'(\Psi^{n-1}(x))},\quad x\in(0,1)
\end{equation}
suffers a jump in $\frac{(1-a)b}{(1-b)a}>1$ times at the point~$a$.
Hence, the same should be true for~$G$ given by~\eqref{l1}.
But this is impossible as $G$ should be concave.

In this example, the non-existence of $(\Psi_t)_{t\ge0}$ is brought by the
discontinuity of~$\Psi'$.
However, one can modify this example by taking $b$ close enough to~1 and by
smoothing~$\Psi$ in some small neighborhood~$U$ of~$a$.
Then $\Psi'(x)$ would suffer a decrease in $\frac{(1-a)b}{(1-b)a}>1$ times
over~$U$.
At the same time the functions $\Psi'(\Psi^n(x))$ with $n\ge2$ are constant
in~$U$, while $\Psi^n(x)-1=c_n(\Psi(x)-1)$ in~$U$ with some constant~$c_n$.
If $U$ is small enough, then the relative change of the latter function
over~$U$ does not exceed~2, so that the relative change of each~$G_n$ given
by~\eqref{l2} over~$U$ suffers a decrease in at least
$\frac{(1-a)b}{2(1-b)a}>1$ times over~$U$.
The appropriate choice of~$a$, $b$, and~$U$ would then contradict the
concavity of~$G$.~\hfill$\Box$
\end{example}

\section{Operations on Distortion Semigroups}
\label{O}

The class of generators of distortion semigroups, i.e. concave functions
$(0,1)\to(0,1)$ is closed under several operations (e.g., the minimum of two
generators is again a generator). It is interesting to see what the
corresponding operations are on the corresponding semigroups
or coherent indices.

\bigskip
\textbf{1. Scaling.}
Let $G$ be a generator, $(\Psi_t)_{t\ge0}$ be the corresponding semigroup
given by~\eqref{r1}, and $\alpha$ be the corresponding index given
by~\eqref{i2}. Let $\lambda$ be a positive number.
Then $\widetilde G=\lambda G$ is a again a generator.
The corresponding semigroup has the form
$\widetilde\Psi_t=\Psi_{\lambda t}$ and the corresponding index is
$\widetilde\alpha(X)=\lambda^{-1}\alpha(X)$.

Let us remark that if $\alpha$ is an arbitrary coherent index and
$\varphi:\mathbb{R}_+\to\mathbb{R}_+$ is a strictly increasing continuous function,
then $\widetilde\alpha=\varphi\circ\alpha$ is again a coherent index.
However, if both $\alpha$ and $\widetilde\alpha$ are semigroup indices, i.e.
they correspond to distortion semigroups, then $\varphi$ should be linear
(indeed, the corresponding families of concave distortions should be related
by $\widetilde\Psi_t=\Psi_{\varphi^{-1}(t)}$, and then the generator of
$(\widetilde\Psi_t)_{t\ge0}$ should be a multiple of the generator
of~$(\Psi_t)_{t\ge0}$).

\bigskip
\textbf{2. Duality.}
Let $G$ be a generator and $(\Psi_t)_{t\ge0}$ be the corresponding semigroup.
It is easy to see that then the semigroup corresponding to
$\widetilde G(x)=G(1-x)$ has the form $\widetilde\Psi_t(x)=1-\Psi_t^{-1}(1-x)$,
where $\Psi_t^{-1}(x)=\inf\{y:\Psi_t(y)>x\}$, $\inf\emptyset=1$.
In other words, the graph of $\widetilde\Psi_t$ is the reflection of the graph
of~$\Psi_t$ with respect to the axis $y=1-x$.
The semigroup $(\widetilde\Psi_t)_{t\ge0}$ might be called the \textit{dual} to
the semigroup $(\Psi_t)_{t\ge0}$.

As an example, the semigroups given by~\eqref{i4} and~\eqref{i5} are duals to each other.
The dual to semigroup~\eqref{i3} is given by
$$
\Psi_t(x)=e^{-t}x+1-e^{-t},\quad t\ge0,\;x\in(0,1].
$$
Then
$$
\int_\mathbb{R}xd(\Psi_t(D_X(x)))=(1-e^{-t})\mathop{\rm essinf}_\omega X(\omega)
+e^{-t}\mathsf{E}X.
$$
This semigroup does not have nice properties as $\Psi_t(0+)>0$.
Finally, the semigroup~\eqref{i6} is self-adjoint in the sense that its dual
coincides with itself.

\bigskip
\textbf{3. Mixture.}
Let $G^i$ be generators and $(\Psi_t^i)_{t\ge0}$ be the corresponding
semigroups, $i=1,2$.
Then $G=G^1+G^2$ is again a generator.
It follows from Theorem~\ref{R1} combined with Lemma~\ref{R2} that
the corresponding semigroup satisfies
$$
\Psi_t(x)=\lim_{n\to\infty}
\bigl(\Psi_{\frac{t}{n}}^1\circ\Psi_{\frac{t}{n}}^2\bigr)^n,
\quad t\ge0,\;x\in(0,1].
$$
This operation might be called the \textit{mixture} of semigroups.

\bigskip
\textbf{4. Minimum.}
Let $G^i$ be generators, $(\Psi_t^i)_{t\ge0}$ be the corresponding semigroups,
and $\alpha^i$ be the corresponding indices, $i=1,2$.
Then $G=G^1\wedge G^2$ is again a generator and the corresponding semigroup
$(\Psi_t)_{t\ge0}$ is the maximal distortion semigroup dominated by both
$(\Psi_t^1)_{t\ge0}$ and $(\Psi_t^2)_{t\ge0}$.
Indeed, if a semigroup $(\widetilde\Psi_t)_{t\ge0}$ is dominated by both
$(\Psi_t^1)_{t\ge0}$ and $(\Psi_t^2)_{t\ge0}$, then its generator
$\widetilde G$ is dominated by both~$G^1$ and~$G^2$.

The corresponding index~$\alpha$ is then characterized as the smallest
semigroup index (i.e. an index corresponding to a distortion semigroup) that
dominates both~$\alpha^1$ and~$\alpha^2$.
This is seen from the following property, which is easy to prove:
a semigroup index $\widetilde\alpha$ corresponding to a semigroup
$(\widetilde\Psi_t)_{t\ge0}$ dominates a semigroup index
$\overline\alpha$ corresponding to a semigroup $(\overline\Psi_t)_{t\ge0}$ if and
only if $\widetilde\Psi_t\le\overline\Psi_t$ for any~$t\ge0$.

\bigskip
\textbf{5. Maximum.}
Let $G^i$ be generators, $(\Psi_t^i)_{t\ge0}$ be the corresponding
semigroups, and $\alpha^i$ be the corresponding indices, $i=1,2$.
Let $G$ be the smallest concave majorant of~$G^1$ and~$G^2$.
Then $G$ is again a generator and the corresponding semigroup
$(\Psi_t)_{t\ge0}$ is the minimal distortion semigroup dominating both
$(\Psi_t^1)_{t\ge0}$ and $(\Psi_t^2)_{t\ge0}$.

The corresponding index~$\alpha$ is then characterized as the largest
semigroup index that is dominated by both~$\alpha^1$ and~$\alpha^2$.

\section{Conclusion}
\label{C}

The aim of this paper is to extract a sufficiently small subclass
from the class of coherent acceptability indices introduced in~\cite{CM071}.
For this, we have considered the indices from the AIW class
corresponding to a family of concave distortions
$(\Psi_t)_{t\ge0}$ satisfying the semigroup property
$$
\Psi_t\circ\Psi_s=\Psi_{s+t},\quad s,t\ge0.
$$
These are called \textit{concave distortion semigroups}.

It has been proved that distortion semigroups are in a one to one
correspondence with concave functions $G:(0,1)\to(0,\infty)$ via the relations
\begin{align*}
\Psi_t(x)&=\inf\Bigl\{y\in[x,1]:\int_x^y\frac{1}{G(s)}\,ds=t\Bigr\},
\quad t\ge0,\;x\in(0,1],\\
G(x)&=\lim_{t\downarrow0}\frac{\Psi_t(x)-x}{t},\quad x\in(0,1).
\end{align*}
The function $G$ is called the \textit{generator} of the semigroup
$(\Psi_t)_{t\ge0}$.

In order to further narrow the class of good distortion semigroups,
we have considered several properties desirable from the mathematical
or the economical perspective. Examples of semigroups having all
the best properties are provided by the AIMAX semigroup and the Wang transform.
Table~2 gives a corresponding overview of performance measures in decreasing order of generality.
The above description of all distortion semigroups, combined with the discussions
relating their properties with generators, allows one to construct further nice
representatives of distortion semigroups, which in some cases might be more convenient
than~\eqref{i5} and~\eqref{i6}.

We have also considered the problem of finding a semigroup
$(\Psi_t)_{t\ge0}$ with $\Psi_1=\Psi$, where $\Psi$ is a given
concave distortion. This problem has potential practical
applications because by comparing the physical measure (extracted
from the data) and the risk-neutral measure (extracted from option
prices), one can derive the concave distortion $\Psi_{t_*}$ for
some value~$t_*$, and then the problem of recovering the whole
semigroup $(\Psi_t)_{t\ge0}$ arises.
We have proved the uniqueness of the solution to this problem
under the additional assumption $\Psi'_-(1)>0$.
An approximation algorithm for finding the corresponding generator has also
been provided.
It was also shown that such a semigroup does not exist for all concave distortion
functions~$\Psi$.
The problem of describing the class of functions~$\Psi$, for which
there exists such a semigroup, remains open.

The class of coherent acceptability indices corresponding to distortion semigroups
is closed under several interesting operations.

\vspace{3mm}
\begin{center}
\begin{tabular}{|l||l|}
\hline
\rule{0mm}{6mm}\textbf{Class of performance measures}&
\textbf{Examples}\\[1mm]
\hhline{|=||=|}
\rule{0mm}{5mm}General performance measures&
$\text{SR}(X)=\mathsf{E}X/\sigma(X)$,\\[1mm]
&$\text{RAROC}(X)=\mathsf{E}X/\text{V@R}(X)$\\[1mm]
\hline
\rule{0mm}{5mm}Coherent acceptability indices&
$\text{GLR}(X)=\mathsf{E}X/\mathsf{E}X^-$\\[1mm]
\hline
\rule{0mm}{5mm}AIW indices&
$\text{CRAROC}(X)=\mathsf{E}X/\rho(X)$\\[1mm]
&with $\rho$ from the WV@R class\\[1mm]
\hline
\rule{0mm}{5mm}Semigroup indices&
\eqref{i2}+\eqref{i3}, \eqref{i2}+\eqref{i4}\\[1mm]
\hline
\rule{0mm}{5mm}Semigroup indices with nice properties&
\eqref{i2}+\eqref{i5}, \eqref{i2}+\eqref{i6}\\[1mm]
\hline
\end{tabular}

\nopagebreak
\vspace{5mm}
\parbox{104mm}{\small\textbf{Table~2.} The left column displays
five classes of performance measures given in decreasing order of generality.
For each of these classes, the right column provides examples of performance
measures belonging to this class and not to the next~one.}
\end{center}



\newpage
\begin{thebibliography}{100}

\bibitem{A02} \textit{C.~Acerbi.}
Spectral measures of risk: coherent representation of subjective risk aversion.
Journal of Banking and Finance, {\bf 26} (2002), No.~7, p.~1505--1518.

\bibitem{ADEH97} \textit{P.~Artzner, F.~Delbaen, J.-M.~Eber, D.~Heath.}
Thinking coherently.
Risk, {\bf 10} (1997), No.~11, p.~68--71.

\bibitem{ADEH99} \textit{P.~Artzner, F.~Delbaen, J.-M.~Eber, D.~Heath.}
Coherent measures of risk.
Mathematical Finance, {\bf 9} (1999), No.~3, p.~203--228.

\bibitem{C06} \textit{A.S.~Cherny.}
Weighted V@R and its properties.
Finance and Stochastics, {\bf 10} (2006), No.~3, p.~367--393.

\bibitem{C071} \textit{A.S.~Cherny.}
Pricing with coherent risk.
Theory of Probability and Its Applications, {\bf 52} (2007), No.~3, 30~p.

\bibitem{CM071} \textit{A.S.~Cherny, D.B.~Madan.}
On measuring the degree of market efficiency.
Working paper, available at: \texttt{www.ssrn.com}.

\bibitem{CM073} \textit{A.S.~Cherny, D.B.~Madan.}
Recovering coherent risk aversion and scaling consistency.
Working paper.

\bibitem{D90} \textit{D.~Denneberg.}
Distorted probabilities and insurance premium.
Methods of Operations Research, {\bf 52} (1990), p.~21--42.

\bibitem{D94} \textit{B.~Dupire.}
Pricing with a smile.
Risk, {\bf 7} (1994), No.~1, p.~18--20.

\bibitem{EM07} \textit{E.~Eberlein, D.~Madan.}
From required returns to required Sharpe ratios.
Working paper.

\bibitem{FS04} \textit{H.~F\"ollmer, A.~Schied.}
Stochastic finance. An introduction in discrete time.
2nd Ed., Walter de Gruyter, 2004.

\bibitem{IRW01} \textit{R.B.~Israel, J.S.~Rosenthal, J.Z.~Wei.}
Finding generators for Markov chains via empirical transition matrices, with
applications to credit ratings.
Mathematical Finance, {\bf 11} (2001), p.~245--265.

\bibitem{W00} \textit{S.~Wang.}
A class of distortion operators for pricing financial and insurance risks.
Journal of Risk and Insurance, {\bf 67} (2000), p.~15--36.

\bibitem{Y87} \textit{M.E.~Yaari.}
The dual theory of choice under risk.
Econometrica, {\bf 55} (1987), p.~95--115.

\end{thebibliography}
\end{document}